\documentstyle[twoside,fleqn,espcrc2,epsf]{article}

\newcommand{\AmS}{{\protect\the\textfont2
  A\kern-.1667em\lower.5ex\hbox{M}\kern-.125emS}}

\title{  Chiral condensate, quark charge and chiral density 
         \thanks{Supported in part by FWF under Contract No. P11456} 
}

\author{Harald Markum, Wolfgang Sakuler  and Stefan Thurner \\
\vspace{3mm}
Institut f\"{u}r Kernphysik, TU Wien,
         Wiedner Hauptstra\ss e 8-10, A-1040 Vienna, Austria\\ 
}

\begin{document}

\begin{abstract}
We study  the topological and fermionic vacuum structure 
of four-dimensional QCD on the lattice by means of correlators of 
fermionic observables and topological densities. 
We show the existence of strong  
local correlations between the  topological charge
density and the quark condensate, charge and chiral density. 
By  analysis of individual 
gauge configurations, we visualize that instantons (antiinstantons)  
carry positive (negative) chirality, whereas the quark charge density 
fluctuates in sign within instantons. 
\end{abstract}

\maketitle


Over the last two decades several models have been developed to 
describe the basic properties of QCD, namely quark confinement 
and chiral symmetry breaking. The most popular are 
the dual superconductor, leading to confinement, and 
the instanton liquid model, which explains chiral symmetry breaking 
and solves the $U_A(1)$ problem \cite{SHU88}. Both  models 
rely  on the existence of topological excitations, 
monopoles and instantons.  
Instantons have integer topological charge $Q$ which is related to  
the zero eigenvalues of the fermionic matrix 
with  a gauge field configuration 
via the Atiyah-Singer index theorem \cite{singer}.
Apart from this famous connection of topology and fermionic 
degrees of freedom, here we attempt to systematically shed light 
on the relationship between the sea-quark distribution 
and topology \cite{warsch96}. 
We do this by studying correlators of topological densities  
with fermionic observables  of the form 
$ \bar\psi\Gamma\psi$ with $\Gamma = 1,\gamma_4,\gamma_5$.
Those quantities are usually referred to as the  
quark condensate, quark charge density, and the chiral density. 

For the implementation of the topological charge on a Euclidian lattice
we restrict ourselves to the so-called field theoretic definitions which
approximate the topological charge density in the continuum,
$
q(x)=\frac{g^{2}}{32\pi^{2}} \epsilon^{\mu\nu\rho\sigma}
\ \mbox{\rm Tr} \ \Big ( F_{\mu\nu}(x) F_{\rho\sigma}(x) \Big ) \ .
$
We used the plaquette and the hypercube prescription.
To get rid of  quantum fluctuations and 
renormalization constants,
we employed the Cabibbo-Marinari cooling method.
Mathematically and numerically  
the local quark condensate $\bar \psi \psi (x)$ 
is a diagonal element of the inverse of the fermionic matrix
of the QCD action. The other fermionic operators 
are obtained by inserting the Euclidian $\gamma_4$ 
and $\gamma_5$ matrices. 
We compute correlation functions between two observables ${\cal O}_1(x)$ and ${\cal O}_2(y)$ 
\begin{equation}
\label{correlations}
g(y-x)=\langle {\cal O}_1(x) {\cal O}_2(y) \rangle - 
       \langle {\cal O}_1\rangle \langle {\cal O}_2\rangle
\end{equation}
and normalize them to the smallest lattice 
separation $d_{\rm min}$, 
$ c(y-x)=g(y-x)/g(d_{\rm min})$. 
Since  topological objects with opposite sign are equally distributed,
we correlate the  
quark condensate with the square of the topological charge density,
and similarly for the other quantities.

Our simulations were performed for full SU(3) QCD on an 
$8^{3} \times 4$ lattice with
periodic boundary conditions.  Applying a standard Metropolis algorithm 
has the advantage that tunneling between sectors of different topological 
charges occurs at reasonable rates. 
Dynamical quarks in Kogut-Susskind discretization   
with $n_f=3$ flavors of degenerate  mass $m=0.1$ were taken into account using the 
pseudofermionic method. 
We performed runs  in the confinement phase at $\beta=5.2$. 
Measurements were taken on 2000 configurations separated 
by 50 sweeps.

\begin{figure}
\begin{center}
\begin{tabular}{c}
\hspace*{-7mm} \epsfxsize=8.0cm\epsffile{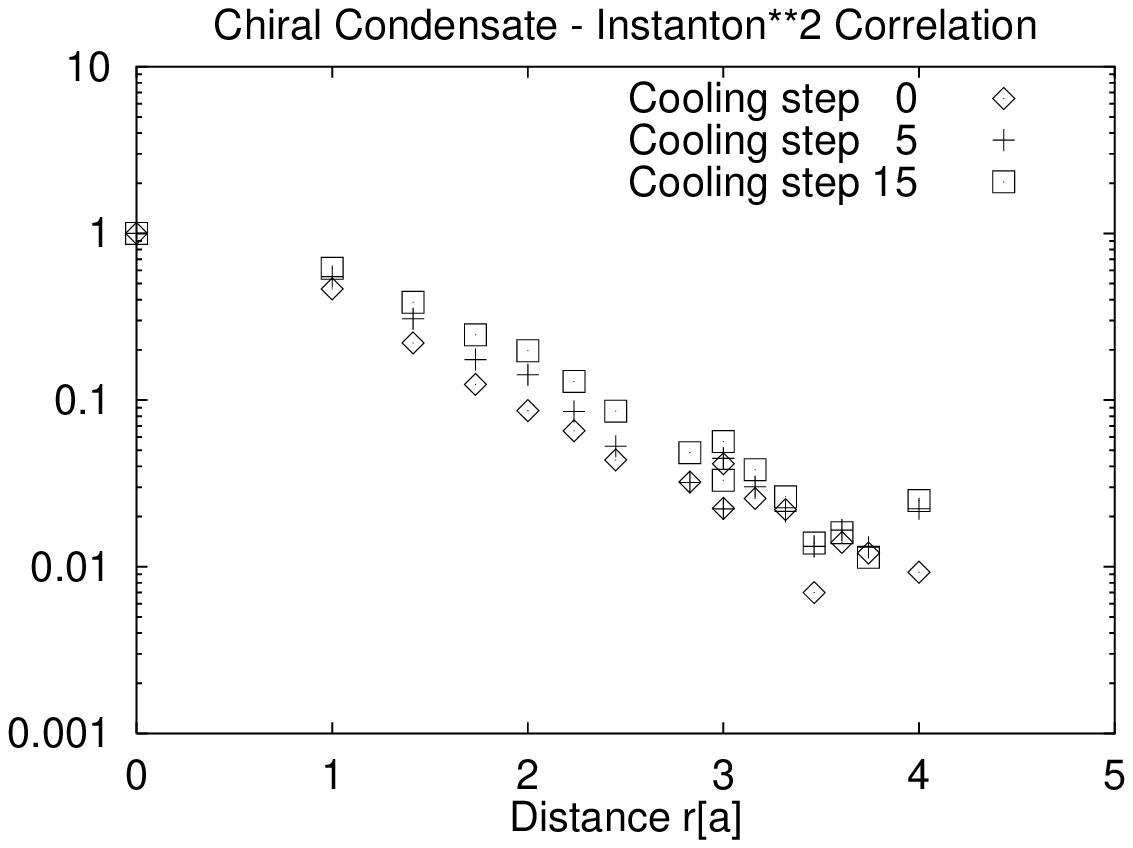} \\[0.3cm]
\hspace*{-7mm} \epsfxsize=8.0cm\epsffile{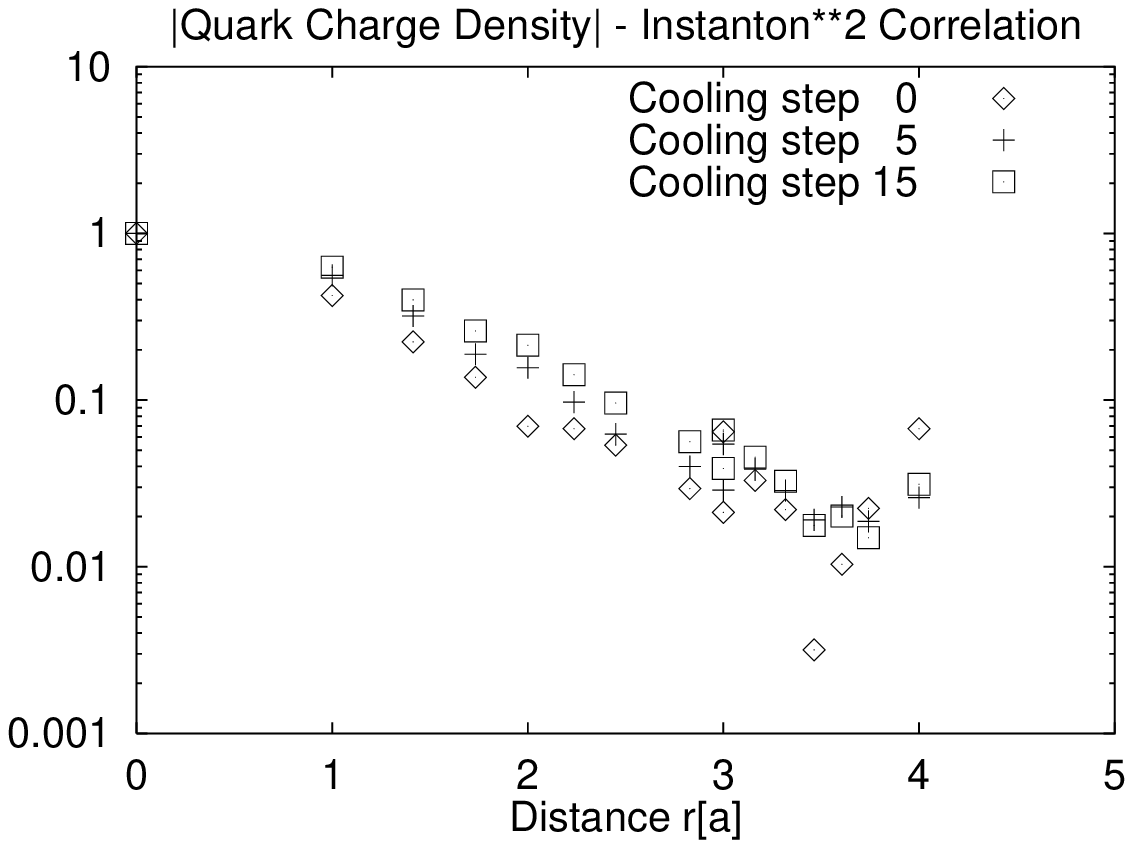} \\[0.3cm]
\hspace*{-7mm} \epsfxsize=8.0cm\epsffile{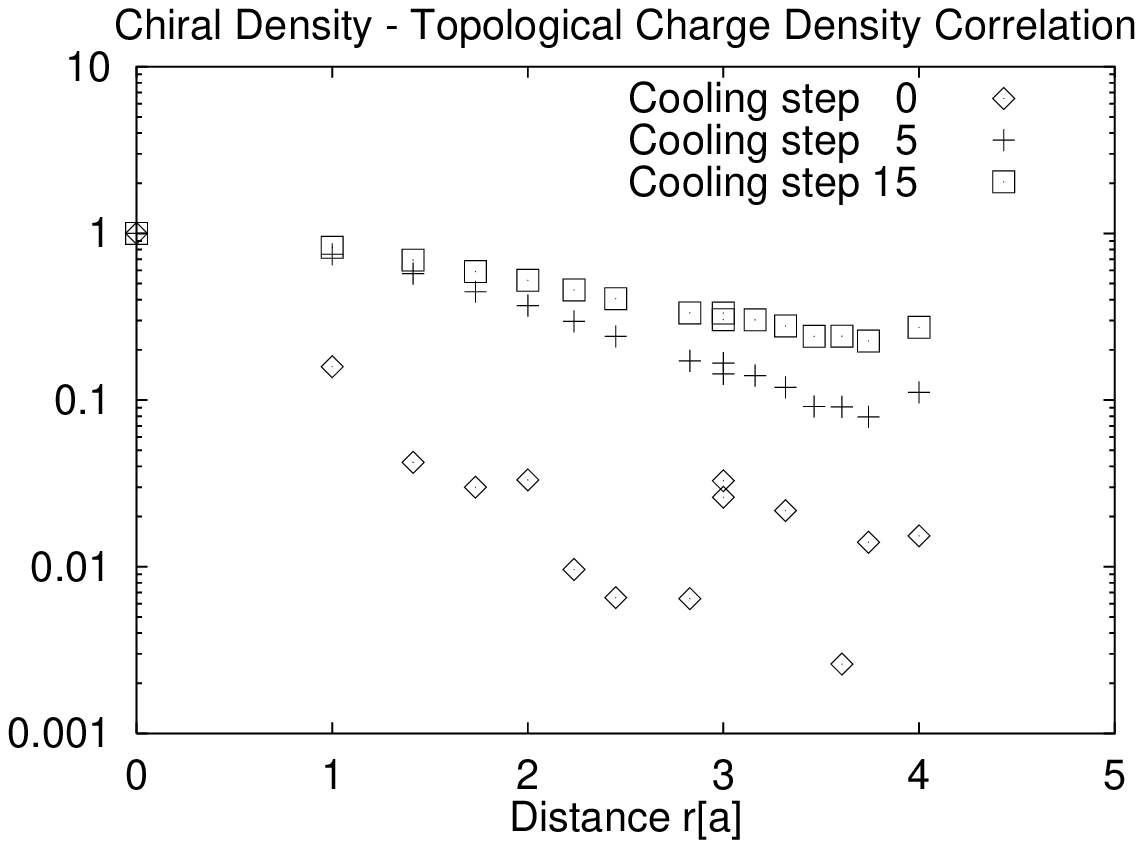}
\end{tabular}
\end{center}

\vspace{-9mm}
\caption{
Correlation functions of topological charge density
with the quark condensate, quark charge
and chiral density shown in the first, second and third
row, respectively.
}
\label{corr}
\end{figure}

Figure~\ref{corr} shows results for the correlation functions 
of Eq.~(\ref{correlations}) with 
${\cal O}_1$ a local fermionic observable and $ {\cal O}_2$
 a topological density.  
All correlations exhibit an extension  
of several lattice spacings. 
Although cooling of quantum fields is necessary to extract topological 
structure, the correlations between the topological charge density and 
both the local chiral condensate and the absolute value of the quark charge density are nearly cooling-independent. 
However, cooling (or some other kind of smoothing) 
is inevitable to obtain nontrivial correlations between the chiral density, 
${\cal O}_1=\bar \psi \gamma_5 \psi(x)$, and the topological charge density. This can be expected since both quantities are correlated via the anomaly. 

We now turn to a direct visualization of  fermionic densities   and
topological quantities on individual gauge fields rather than 
performing gauge averages. We persue this in the following to 
get insight into the local interplay of topology with   
the sea-quark distribution.  
By analyzing dozens of gluon and quark field configurations we obtained the
following results.
The topological charge is hidden in  quantum fluctuations and
becomes visible by cooling of the gauge fields. For
0 cooling steps no structure can be seen in $q(x)$, the fermionic 
observables 
or the monopole currents, which does not mean the absence of correlations
between them. After a few  cooling steps clusters of
 nonzero topological charge density and quark  fields  are resolved.
For more cooling steps both topological charge and
quark fields  begin to die out and eventually vanish.

\begin{figure}
\vspace{-17mm}
\begin{center}
\begin{tabular}{c}
\epsfxsize=6.0cm\epsffile{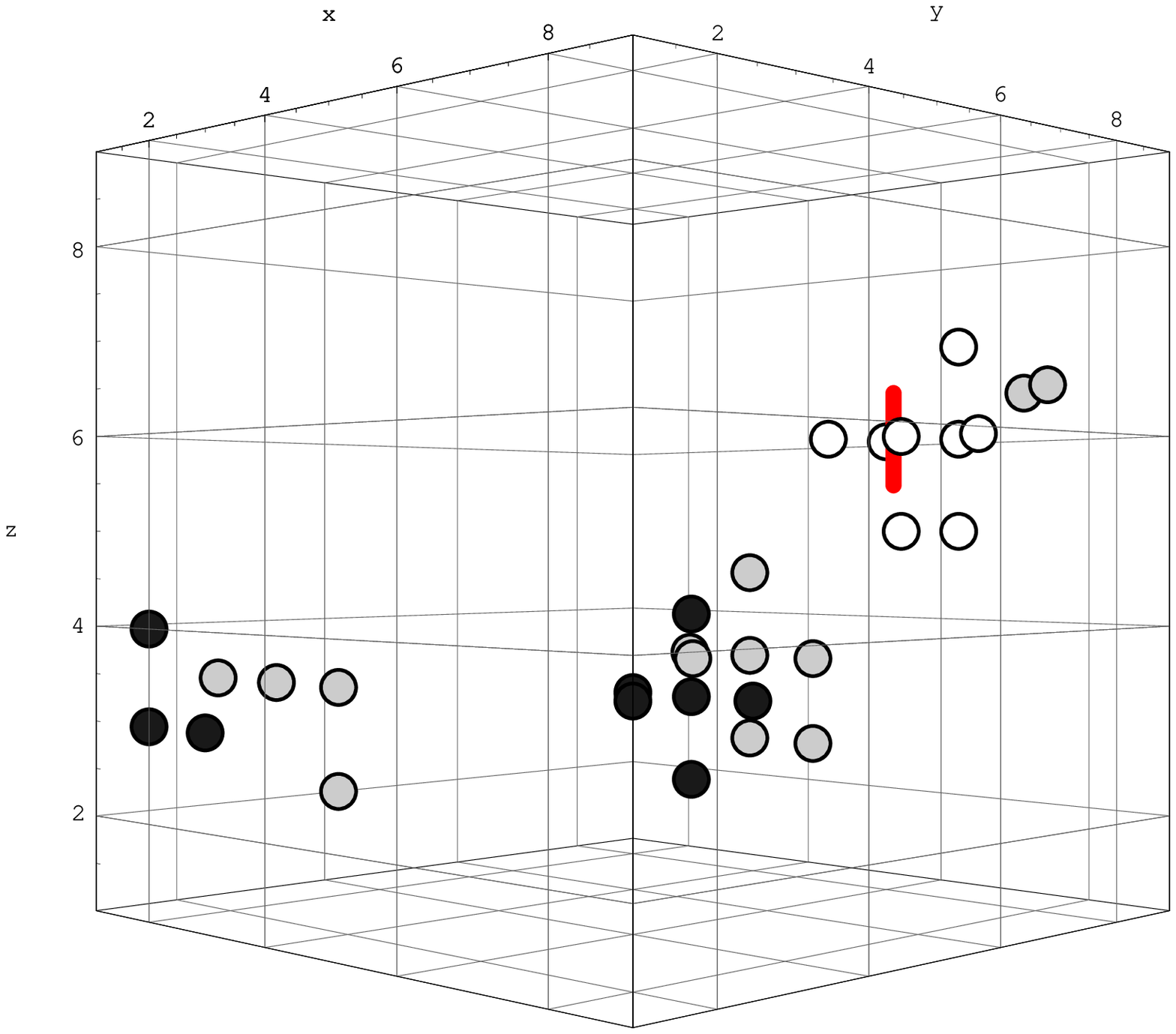 }
\end{tabular}
\end{center}
\vspace{-37mm}
\begin{center}
\begin{tabular}{c}
\epsfxsize=6.0cm\epsffile{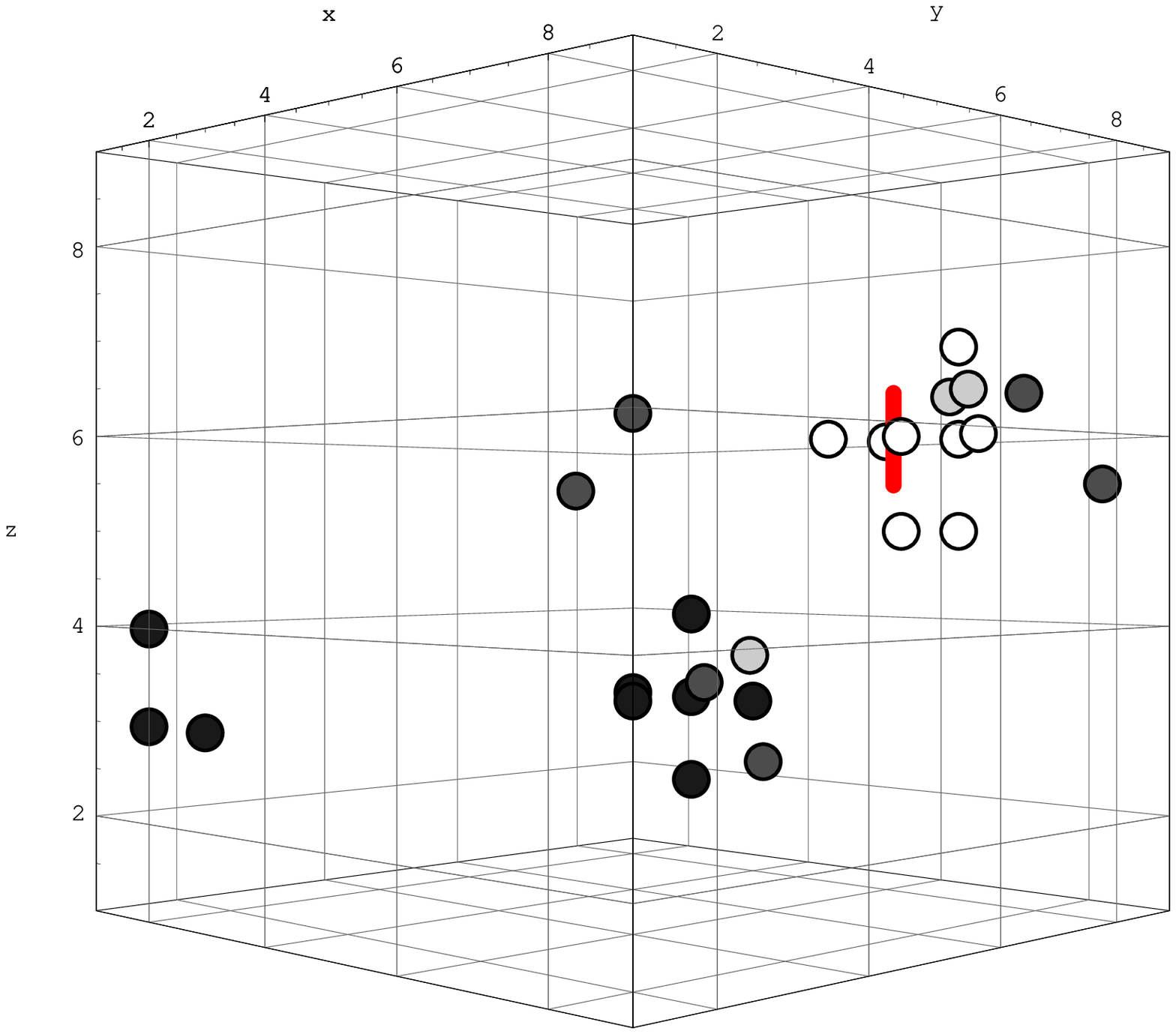 }
\end{tabular}
\end{center}
\vspace{-37mm}
\begin{center}
\begin{tabular}{c}
\epsfxsize=6.0cm\epsffile{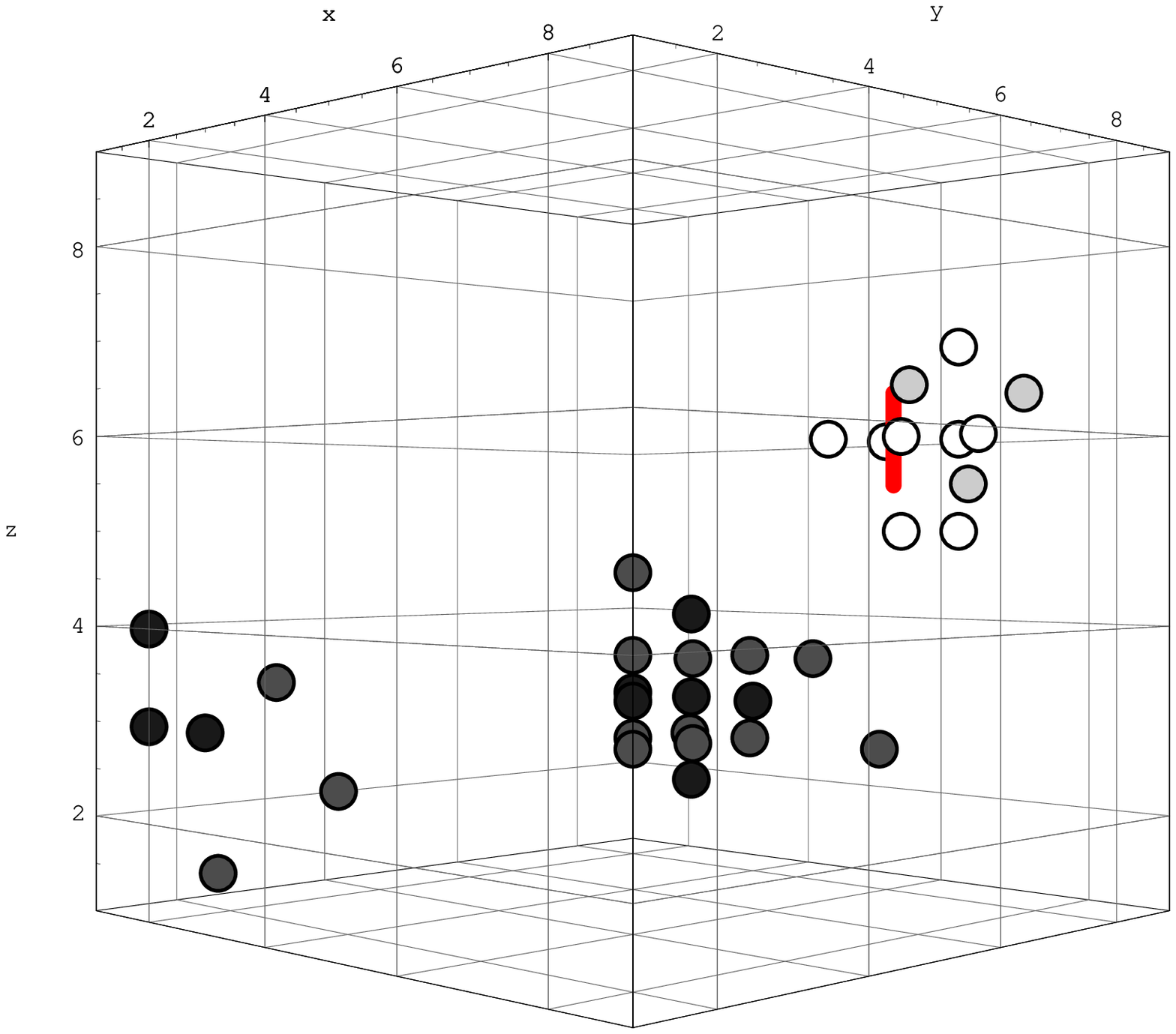 }
\end{tabular}
\end{center}
\vspace{-24mm}

\caption{
Aspects of the fermionic field for a fixed time slice of a gauge field
with an instanton-antiinstanton pair drawn by white and black dots.
Upper plot:
$\bar\psi\psi(x) > 0$ (light grey dots) within the instantons.
Middle plot:
$\psi^{\dag}\psi(x)$ alternates in sign (light and dark grey dots) in a single instanton.
Lower plot:  $\bar \psi\gamma_5\psi(x)$ is positive/negative in instanton/antiinstanton.}
\label{hist}
\end{figure}

In Fig.~\ref{hist}    
a  typical topologically nontrivial 
configuration, consisting of an instanton and an antiinstanton, 
 from SU(3) theory
with dynamical quarks on the $8^{3} \times 4$ lattice 
in the confinement phase is shown after 15 cooling steps 
for fixed time slices.
We display the positive/negative topological charge density  by  
white/black dots if the absolute value exceeds certain minimal fluctuations. 
Monopole currents are defined in the maximum Abelian projection 
and only one type is shown  by lines. 
The upper 3D plot includes the local chiral condensate $\bar\psi\psi(x)$, 
indicated by light grey  dots whenever 
a certain threshold is exceeded. One clearly sees that both the instantons and 
antiinstantons are surrounded by a cloud of $\bar\psi\psi(x) > 0$ \cite{HANDS}. The middle 
3D plot exhibits the situation for the quark charge density 
$\psi^{\dag}\psi(x)$ indicated by light and dark grey dots depending on the 
sign of the net color charge excess. One observes that $\psi^{\dag}\psi(x)$ 
alternates in sign already in one instanton implying trivial 
correlations $<\psi^{\dag}\psi(x) q(y)>=0$ (not shown in Fig. 1). 
The lower 3D plot displays the chiral density $\bar \psi \gamma_5 \psi(x)$ 
again indicated by light and dark grey dots. One nicely sees that the positive instanton 
is always surrounded by a lump with $\bar \psi \gamma_5 \psi(x)>0$ and vice versa.
Combining the above finding of  Fig.~1 showing that the correlation 
functions between fermionic  and topological  quantities  
are not very sensitive to cooling together with  the 
3D images in Fig.~2, we conclude that instantons go hand in hand with clusters of 
$\bar \psi\Gamma \psi (x)\neq 0$ , $\Gamma=1,\gamma_4,\gamma_5$, 
also in the uncooled QCD vacuum.  


In summary, our calculations of correlation functions
between topological densities and the fermionic observables yield an
exponential decrease. 
Results for the condensate and the modulus of the quark charge correlators are almost 
identical, as expected, since the quark condensate 
reflects the absolute value of the quark charge density. These 
correlation functions show little cooling dependence. 

The correlations unambiguously demonstrate that not only the local 
chiral condensate but also the quark charge and chiral density    
take non-vanishing values predominantly in the regions  of
instantons and monopole loops.
Note that for the chiral density this behavior is expected due to the 
anomaly. 

Visualization exhibited that the distribution of 
sea-quarks is drastically enhanced around centers of nontrivial 
topology (instantons, monopoles) in Euclidian space-time. 
It must be emphasized that this represents the situation on a finite lattice with 
finite quark mass without the extrapolation  to the thermodynamic and chiral limit. 
However, since all such correlators turned out rather independent of the 
gauge group and choice of the action etc., we expect that they are generic.

%


\end{document}